%% file: main.tex
\documentclass[11pt]{article}

\usepackage{authblk}

\usepackage[english]{babel}
\usepackage[utf8]{inputenc}
\usepackage[T1]{fontenc}
\usepackage{lmodern}

\usepackage[dvipsnames]{xcolor}
\usepackage[misc]{ifsym}
\usepackage{soul}

\usepackage[letterpaper,left=3cm,right=3cm,top=3cm,bottom=3cm]{geometry}

\usepackage{amsmath,amsfonts,amsthm,amssymb,mathrsfs,dsfont} % Math packages
\usepackage{hyperref}
\hypersetup{
    colorlinks = true,
    allcolors = blue
}
\usepackage[capitalize,nameinlink,noabbrev]{cleveref}
\usepackage{csquotes}
\usepackage{mathtools}
\usepackage{graphicx}
\usepackage{enumerate}
\usepackage[font=small,labelfont=bf, width=0.9\textwidth]{caption}
\usepackage{float}

\usepackage{titlesec}
\titleformat{\section}{\normalfont\Large\bfseries}{\thesection}{10pt}{}
\titleformat{\subsection}{\normalfont\large\bfseries}{\thesubsection}{10pt}{}
\titleformat{\subsubsection}{\normalfont\normalsize\bfseries}{}{0pt}{}

\usepackage[numbers]{natbib}
\usepackage{enumitem}
\setlist[enumerate]{itemsep=0mm}

\definecolor{seabornBlue}{RGB}{76,114,176}
\definecolor{seabornGreen}{RGB}{85,168,104}
\definecolor{seabornRed}{RGB}{196,78,82}

\definecolor{orangePumpkin}{RGB}{211,84,0}
\definecolor{nephritis}{RGB}{39,174,96}
\definecolor{orangeCarrot}{RGB}{230,126,34}
\definecolor{blueBelizeHole}{RGB}{41,128,185}
\definecolor{redAlizarin}{RGB}{231,76,60}
\definecolor{redNasturcianFlower}{RGB}{232,65,24}

\input{macros}

\DeclareMathOperator{\asinh}{asinh}

\newcommand{\knv}[1]{\ensuremath \mathbf{v}}

\title{\bf Comment on ``mbtransfer: Microbiome intervention analysis using transfer functions and mirror statistics'': Implementation errors, theoretical misapplication, and methodological flaws}

\author[1,2,3,4,\Letter]{Travis E. Gibson}

\affil[1]{Harvard Medical School, Boston, MA USA}
\affil[2]{Brigham and Women's Hospital, Boston MA, USA}
\affil[3]{Broad Institute of MIT and Harvard, Cambridge, MA, USA}
\affil[4]{Massachusetts Institute of Technology, Cambridge, MA, USA}

\affil[ \Letter ]{\footnotesize\href{mailto:tegibson@bwh.harvard.edu}{\texttt{tegibson@bwh.harvard.edu}}}

\usepackage{changepage}
\usepackage{setspace}
\makeatletter
\let\oldaffillist\AB@affillist
\renewcommand{\AB@affillist}{\vspace{0.1in}
\begin{adjustwidth}{1in}{1in}\begin{flushleft}
\oldaffillist
\end{flushleft}\end{adjustwidth}}
\makeatother

\date{}

\usepackage{sansmath}
\sansmath

\defcitealias{mbt}{MBT}
\def\mbt{\citetalias{mbt}}
\def\mbtransfer{\texttt{mbtransfer}}
\def\mdsine{\texttt{MDSINE2}}

\begin{document}

\maketitle

\begin{abstract}
There are a number of errors in  ``mbtransfer: Microbiome intervention analysis using transfer functions and mirror statistics'' PLOS Comp Bio (2024) spanning multiple aspects of the paper. The wrong inputs were provided to comparator methods for model training, when forecasting one method was provided initial conditions in the wrong units, and  performance metrics were calculated without proper unit conversion. The false discovery rate and power analysis conclusions provided in the text are not supported by theory or the empirical testing that was performed within the paper. The paper also has data leakage issues, equations are written down incorrectly, and incorrect definitions/terminology are used.
\end{abstract}

\section{Introduction}
This critique is regarding  ``mbtransfer: Microbiome intervention analysis using transfer functions and mirror statistics'' PLOS Comp Bio (2024) which we will refer to as \mbt{} for short \cite{mbt}. In \mbt{} a method called \mbtransfer{} is proposed that learns nonlinear dynamics from time-series microbiome data. The class of dynamics it models are linear with respect to microbial abundances, interventions, and fixed subject characteristics, as well as possible (non-linear) interactions between them. The model also explicitly allows for time lagged effects from both interventions and microbial abundances. As part of the analysis presented in \mbt{} the forecasting ability of the model is assessed and the authors also introduce mirror statistics as a means for determining which taxa are significantly perturbed following interventions.  

We wish to point out many critical issues with the analysis in \mbt{} which span all aspects of the original contribution. When performing inference with one of the comparator methods,  \mdsine{} \cite{gibson2025mdsine2}, the authors supplied it the incorrect inputs and ran the model with several orders of magnitude fewer Monte Carlo samples than what was performed in the original work. After \mdsine{} was trained and it was used for forecasting the authors supplied initial conditions in the wrong units and furthermore did not transform the model forecasts to the same units as the hold out data used to calculate performance metrics. These incorrect inputs and unit discrepancies results in data that are many orders of magnitude different from what they should have been. These two issues are discussed in \cref{sec:inference,sec:forecast} respectively.

It is claimed in \mbt{} that mirror statistics provide good False Discovery Rate (FDR) control
with \mbtransfer{} function estimates so long as the number of taxa is large. We will show that the empirical results from their simulations however are not consistent on this point and that the theoretical result they reference to support this claim doesn't actually hold in their setting. This is presented in \cref{sec:fdr}. We will also discuss data leakage issues that occur in \mbt{} (\cref{sec:leakage}) and instances of incorrect equations (for dynamics) and definitions which is presented in \cref{sec:definitions}. We then close with a brief conclusions section.

\section{Inference}\label{sec:inference}
One of the inputs \mdsine{} is expecting is a table of read abundances. For the results in Figure 4 of \mbt{} the method \mdsine{} was not supplied raw read abundances but instead inverse hyperbolic sine (asinh) transformed read abundances ($\asinh=\log(x+\sqrt{x^2+1})$). \mdsine{} has a Negative Binomial noise model for reads so seeing transformed reads with values between 0 and 10 tells the model that the noise is through the roof! This also messes up the proportions of, and changes in, the taxa abundances the model is trying to learn from with regards to generalized Lotka-Volterra (gLV) dynamics (e.g. a microbe doubling in abundance would no longer be reflected as a doubling but whatever the transformed values are that are no longer proportional to the actual changes in the abundances). The authors note that ``[w]hen the data were not asinh transformed, the mbtransfer model performed worse than either MDSINE2 or fido'' and those results are in the supplemental text (\mbt{} S2 Text, Figs E,F).

In the \mbt{} paper they also performed inference with exceedingly fewer Gibbs steps than what was performed in the \mdsine{} paper. In the \mdsine{} paper there were a minimum of 15,000 Gibbs steps (5,000 for burnin and 10,000 posterior) and in some cases we ran up to 10 different MCMC chains for inference and then combined their posterior samples resulting in 100,000 posterior samples. In \mbt{} for benchmarking with \mdsine{} they only performed inference with 225 Gibbs steps (25 for burnin and 200 posterior). That is two orders of magnitude lower than what was used in any of the analysis in the \mdsine{} paper 

\section{Forecasting}\label{sec:forecast}
The forecasting that was performed in \mbt{} with \mdsine{} was also not performed correctly. To understand why, we need to take a moment to explain the difference between AR models and SSMs. With an AR model you are directly modeling the changes over time of the data or measurements you have. If the measurements indexed over time t are denoted as $y_t$ then one possible realization of an AR model could be 
\begin{equation*}
    y_{t+1}=f(y_t)+\epsilon_t
\end{equation*}
Where $f(\cdot)$ captures the transition dynamics and $\epsilon_t$ is a stochastic effect. The important point to emphasize is that the transition dynamics are occurring in the same space as the measurements or data. State space models on the other hand have two spaces of interest, one for the measurements (which we will again denote as $y_t$) and a latent state space where the transition dynamics occur (which we will denote as $x_t$) \cite{durbin2012time}. One possible realization of a SSM could be
\begin{align*}
x_{t+1}&=g(x_t)+\nu_t\\
y_t&=h(x_t)+\omega_t
\end{align*}
where $h(\cdot)$ is a map from the latent states to measurements and $g(\cdot)$ captures the transition dynamics in the latent space with $\nu_t$ and $\omega_t$ stochastics effects capturing the process variance and the measurement noise. The important thing to emphasize here is that $x$ and $y$ can live in different spaces and have different units even. 

\mbtransfer{} is an AR model, if it is supplied read abundances to learn from then, the model can be used to directly forecast read abundances from any initial condition you provide. \mdsine{} is an SSM. It is expecting two different measurement modalities (1) read abundances as we have already discussed, and (2) qPCR triplicate values in terms of total bacterial load for each sample (colony forming units per gram of feces, or CFU/g). The latent states for \mdsine{} are in the units of CFU/g for each individual taxa and the model forecasts microbial abundances in the latent state in those same units. In \mbt{}, instead of supplying an initial condition to simulate from in the units of CFU/g for \mdsine{} they supplied initial conditions for $x$ in the units of reads or nonlinearly transformed reads. This is completely wrong. These values will be many orders of magnitude different from what they should be and so any perceived poor performance will be because \mdsine{} is trying to reach values that are orders of magnitude different than the values it was supplied.\footnote{We manually checked this in the code and the abundances are off by 6 to 7 orders of magnitude. They are supplying initial conditions with a maximum abundance of $\sim$100 while the qPCR values they supplied for training the model was $10^9$.  Our code where we verified this can be found here \url{https://github.com/gibsonlab/microbiome_interventions/blob/main/_gibson_lab_analysis/cross_validation/console_dump.txt}}  

This mismatch in units was also not accounted for properly when computing the performance metrics as well. Even if the model somehow reached an accurate steady state abundance, after having wildly off initial conditions, the forecast that \mdsine{} performed in CFU/g would then need to be converted to the units in which the performance metrics were being evaluated (reads or transformed reads), but this was not done either. The discrepancy is like supplying someone values in miles when they were expecting values in meters. It is worse than that though because the mapping from read counts to CFU/g is not as simple as a linear scaling. This mapping is nonlinear (read depth is not a quantitative measure of actual abundance, and the read depth changes from sample to sample for technical reasons). The more apt comparison is you ask for a distance in meters between two places, and the other party responds with “it takes 2 hours to get there”. In summary, \mdsine{} was both provided the wrong kind of data for training, and for forecasting (and performance metrics computation) was provided values in the wrong units. Thus, even if the forecasting was 100\% accurate there would be significant errors because of the unit mismatch.

There is another, albeit more subtle, issue with how the forecasting with \mdsine{} was performed in \mbt{} as well.
The authors did not properly compute the empirical expected value of the interaction strength variables from the posterior samples that \mdsine{} provides. They included samples (from the prior) for interaction strength variables when the associated indicator sample was zero. That is, the interaction does not exist for those samples. Partly because of the kind of issue encountered here a different forecasting paradigm was adopted in \cite{gibson2025mdsine2}. A forecast was generated for each Gibbs step, generating a distribution for the trajectories, and then the median value of the taxa at each time-point was used as our best estimate. This is discussed in more detail in \cref{app:example}.

\section{False discovery rate (FDR) and power analysis}
\label{sec:fdr}

The simulation results presented in \mbt{} are not consistent with the claims made in the paper and a theoretical result they reference to justify their claims is incorrectly invoked. In \mbt{} it is stated

 \begin{quote}
 ``Nonetheless, for both high and low phylogenetic correlation, mirrors effectively control the FDR when the number of taxa is large and the data have been DESeq2-asinh normalized. This is consistent with the improved forecasting performance for transformed data and with Proposition 3.3 of [\citeauthor{dai2023false} \citeyear{dai2023false}], which guarantees FDR control asymptotically as the number of hypotheses increases.'' - \cite[Pages 11-12]{mbt}
 \end{quote}
The empirical results and theoretical claims in the above quote are addressed in detail within the two subsequent subsections.

\subsection{Inconsistent empirical results}

\begin{figure}[b!]
    \centering
\includegraphics[width=0.75\textwidth]{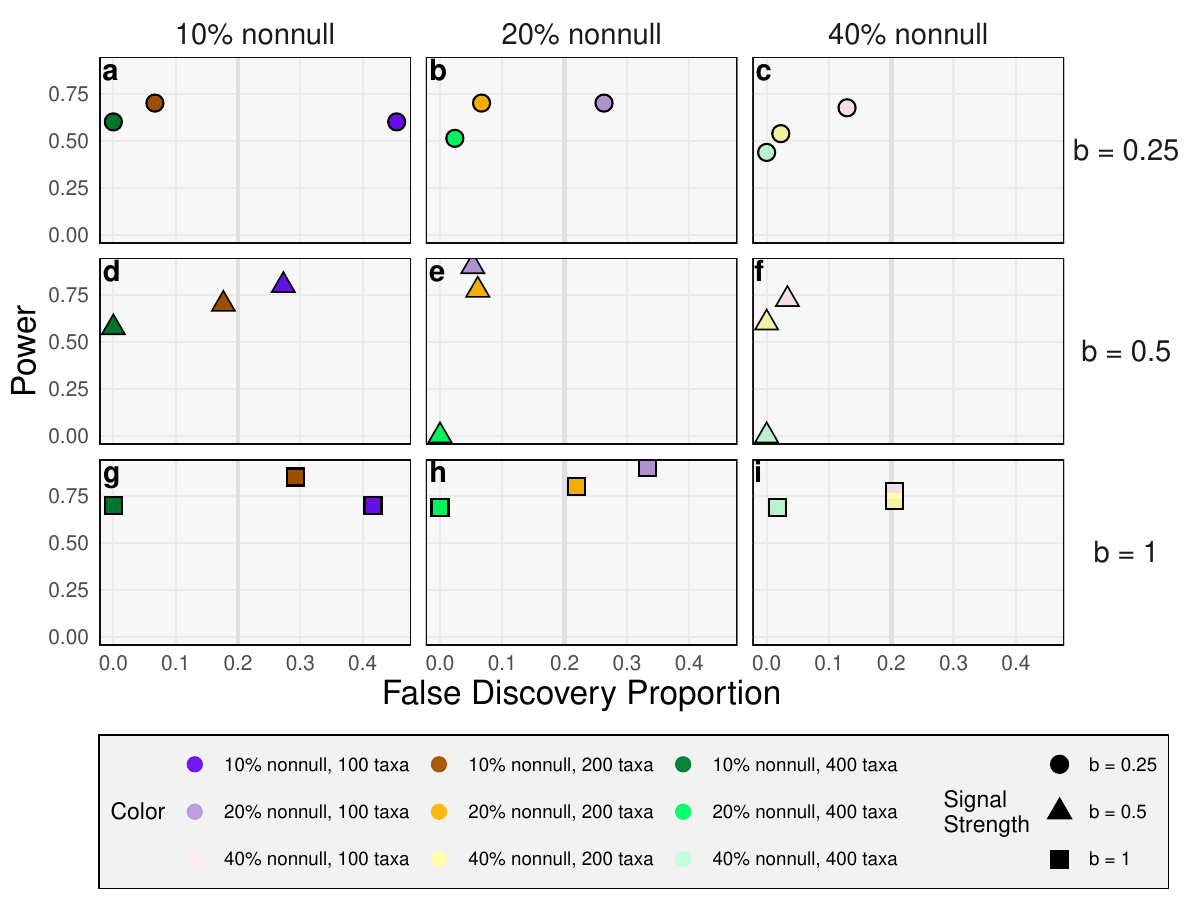}
    \caption{ \textbf{Power vs False Discovery Proportion: varying the number of taxa within each subplot.} DESeq2-asinh normalized reads with $\text{Lag}=1$ and $\alpha=0.1$.
    }\label{fig:1}
\end{figure}

\begin{figure}[b!]
    \centering
\includegraphics[width=0.75\textwidth]{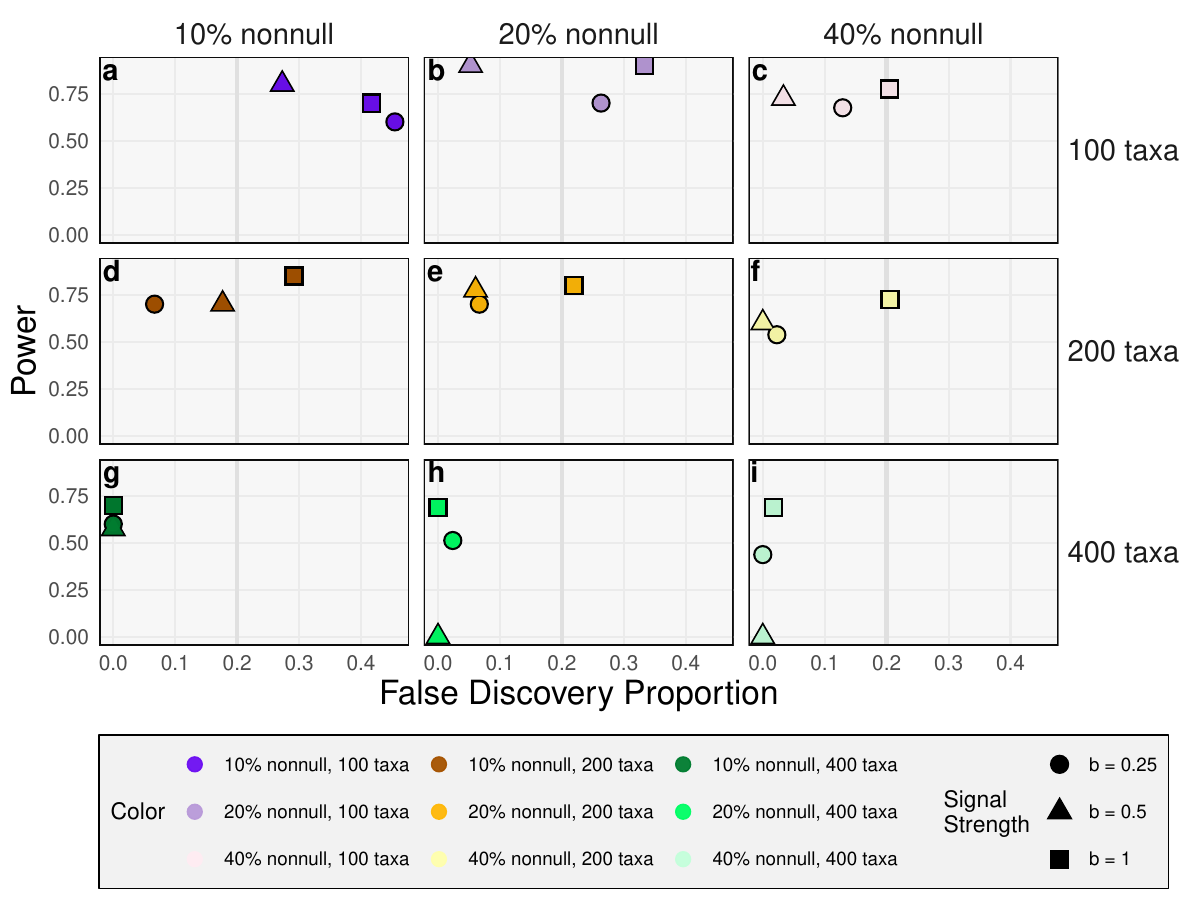}
    \caption{ \textbf{Power vs False Discovery Proportion: varying the signal strength within each subplot.} DESeq2-asinh normalized reads with $\text{Lag}=1$ and $\alpha=0.1$.
    }\label{fig:2}
\end{figure}

It is hard to tell how well mirrors controls the FDR as presented in the original presentation as certain parameters are varied (see original \cite[Figure 6]{mbt}). In order to see how the FDR changes with different simulation variables we generated new plots using their intermediate saved data (downloaded from this address \url{https://go.wisc.edu/3gc982}) where we varied only one simulation variable at a time within each subplot (\cref{fig:1,fig:2}). 

When varying the number of taxa we see an inconsistent relationship between power and FDR. In \cref{fig:1}a we see that by increasing the number of taxa from 100 to 400 the FDP goes from approximately 0.45 to 0 while maintaining high power. However in \cref{fig:1}e the power collapses to zero when their are 400 taxa.

Organizing the same underlying data in a slightly different manner, \cref{fig:2}, we also see very contradictory results when changing the signal strength. In \cref{fig:2}a,b,e we can see that increasing the signal strength causes you to increase the FDP while maintaining or decreasing in power (worse performance). There are however also scenarios where either increasing or decreasing the signal strength improves performance, \cref{fig:2}g,h,i (you might expect with increased signal strength to have this behavior, but not with decreased signal strength).

\clearpage

\subsection{Misinterpretation of Theoretical Guarantees}
The authors incorrectly invoke Proposition 3.3 from \citeauthor{dai2023false} to justify the claim that mirrors "effectively control the FDR when the number of taxa is large" and that Proposition 3.3 "guarantees FDR control asymptotically as the number of hypotheses increases." This statement doesn't accurately characterize the theoretical requirements.

Proposition 3.3 requires {\bf both} the number of hypotheses (the variable $p$ in \citeauthor{dai2023false}) and the number of samples (the variable $n$ in \citeauthor{dai2023false}) to approach infinity at proper rates. The theorem does not guarantee FDR control simply by increasing $p$ while keeping $n$ fixed (as was the case in the simulations performed in \mbt). Furthermore, Proposition 3.3 is about detecting interactions between features (nodes) connected in a network and not about detecting which features are responding to external perturbations. Proposition 3.1 of \citeauthor{dai2023false} is closer in spirit to what is actually being tested for in \mbt. Note that Proposition 3.1  has the same underlying assumption that both $p$ and $n$ approach infinity at appropriate rates.

\section{Data Leakage}\label{sec:leakage}
The \mbt{} paper also has data leakage issues. In many of their results they apply DESeq2\cite{love2014moderated} Normalization in a pre-processing step (for all samples in a bulk process) and include samples which are later used as forecasting test samples when doing so. To see why this is the case let’s review DESeq2 Normalization \cite{love2014moderated}. For reads $K_{ij}$ where $i$ is this taxa index and $j$ is sample index the per sample normalization $s_j$ is performed as follows
\begin{equation*}
s_j = \text{median}_i \left( \frac{K_{ij}}{ \bar K_i} \right)  \quad     \text{where} \quad        \bar K_i=\left(\prod_{j=1}^m K_{ij}\right)^{1/m}
\end{equation*}
The data leakage can occur if hold out samples are included in the per taxa geometric mean calculation $\bar K_i$. In the \mbt{} paper this normalization was applied in a bulk process as we suggested and included samples that would later be used as hold out samples for testing the forecasting performance of the methods. One of the most basic rules in machine learning is to avoid data leakage \cite{gihawi2023major}.

\section{Definitions}\label{sec:definitions}
In \mbt{} gLV dynamics are incorrectly defined. In the text of \mbt{} it is stated: ``[t]he gLV supposes $\partial y(t)/\partial t  =Ay(t)+Dw(t)+\epsilon(t)$''. Those are linear dynamics and not gLV dynamics. 

In \mbt{} it is implied that transfer functions are defined for nonlinear maps. The term transfer function has only ever been used for linear systems because of the many special properties they have \cite{brown1948dp}. mbtransfer learns nonlinear maps for its transition dynamics. The nonlinearity arises from the interactions it learns: ``While components are linear, the xyz-derived interactions [...] induce nonlinearity with respect to the original inputs''. The method mbtransfer is thus named after a representation in dynamical systems theory that, strictly speaking, does not apply to the class of dynamics mbtransfer is modeling.

\section{Conclusions}

In summary, \mdsine{} was provided with: (1) incorrectly transformed input data, (2) $\sim$60-fold fewer MCMC samples than recommended, (3) initial conditions in wrong units differing by 6-7 orders of magnitude, and (4) forecasts evaluated without proper unit conversion. The FDR and power analysis is empirically inconsistent and the theoretical claims are not consistent with the assumptions of the theoretical results they cite. There are also issues with definitions and data leakage as well. The accumulation of errors across all aspects of this work call into question its validity. 

%These findings underscore the importance of methodological diligence when developing and evaluating new statistical methods. 

% ===
% END MAIN TEXT 
% ===

\section*{Code availability}
We forked the original mbtransfer paper repo and added our analysis to it here \url{https://github.com/gibsonlab/microbiome_interventions}. To verify the unit mismatch we performed a console dump which can be found here \url{https://github.com/gibsonlab/microbiome_interventions/blob/main/_gibson_lab_analysis/cross_validation/console_dump.txt}. For creating the FDR and power analysis results in \cref{fig:1,fig:2} we downloaded the ``inference{\_}results'' tar files from here \url{https://go.wisc.edu/3gc982} collated the results and then generated the figures using the following \url{https://github.com/gibsonlab/microbiome_interventions/blob/main/scripts/simulation_interpretation_v2.Rmd}.

\section*{Acknowledgments}
TEG was supported by NIH R35GM143056 and also in part by NSF MTM2 2025512.

\clearpage
\bibliography{sources}

% ======================================= Appendix =========================================
%TC:ignore
\clearpage
\titleformat{\section}{\normalfont\Large\bfseries}{\thesection}{1em}{}
\titleformat{\subsection}{\normalfont\large\bfseries}{\thesubsection}{1em}{}
\titleformat{\subsubsection}{\normalfont\large\bfseries}{\thesubsubsection}{1em}{}

\begin{appendix}

% === Rename "Figure D.1" to "Supplemental Text Figure T1"
\renewcommand{\thefigure}{T\arabic{figure}}
\makeatletter
\renewcommand{\fnum@figure}{Supplemental~Text~Figure~T\arabic{figure}}
\makeatother

\crefalias{section}{appendix}

\begin{center}
    {\Large \textbf{Appendix}\par}
\end{center}

\section{Forecasting and conditional expectations for structural components}
\label{app:example}

 When forecasting with \mdsine{} in our paper \cite{gibson2025mdsine2} we generated a distribution of forecasts and then for each time point we selected the median abundance for each taxa from that distribution as our best guess when computing benchmarks. Specifically, for each Gibbs step in the posterior we take the sampled values for all the parameters in the model, and then with those values we perform a forecast with the model. With this process you end up with as many forecasts are there are Gibbs steps in the posterior of your MCMC chain and then we obtain our best estimate for taxa abundances as just described above. In \mbt{} instead of generating a distribution of forecasts they chose a single set of parameters and then performed one forecast. On its face this could be an acceptable way to forecast. Setting the single best (or even the most likely) value for a parameter isn’t always straightforward though, particularly when a model has structural components. 

In our model the pairwise interactions are modeled with two parameters, an indicator variable and an interaction strength variable. An interaction doesn’t exist when the indicator is zero, so you must be careful when computing the expected value for the interaction strength. For Gibbs samples where an indicator variable has a value of zero, the interaction strength parameter has no effect on the dynamics and its Gibbs sample is simply taken from the prior. The fact that a sample is taken for the interaction strength in this scenario is just so all the MCMC chains are the same length – but strictly speaking an interaction strength doesn’t exist when the indicator is zero. If you want to estimate the most likely interaction strength between taxa you will want to take a conditional expectation (conditioned on the indicator variable being equal to one). In \mbt{} they did not catch this subtle distinction and calculated the expected value of the interactions as the mean of the interaction strength samples over all Gibbs steps which included samples from the prior. This point is further elucidated with a concrete example the following subsection.

\subsection{Toy example with structure learning}

To illustrate the issue raised above we will employ  the following toy example similar in structure to the dynamics of \mdsine{} will be introduced. Consider the simple linear dynamics 
\begin{equation*}
    x_{k+1}=Ax_k     \quad     \left( x_{i,k+1}=\sum a_{ij} x_{j,k} \right)    
\end{equation*}          
Where the $n\times n$ interaction matrix $A$ has a sparse structure with many of its elements being 0 while the other nonzero elements take values in the reals. One way of learning a model for these dynamics that can explicitly account for the sparse structure in the interaction matrix $A$ and that mirrors our modeling in \mdsine{} would be as follows
\begin{equation*}
    x_{k+1}=B\cdot Z x_k     \quad     \left( x_{i,k+1}=\sum b_{ij} z_{ij}x_{j,k} \right)    
\end{equation*}  
where the dot between $B$ and $Z$ denotes the element-wise product. The matrix $B$ takes values in the reals and is meant to capture the strength of interactions. The matrix $Z$ on the other hand only takes values of 1 or 0 and is meant to capture the presence or absence of an interaction. Let’s assume that we have estimates of the posterior distribution of $B$ and $Z$ using MCMC and we have a chain of Gibbs samples. If you wanted to estimate what you think is the most likely interaction parameter (i.e. what is your best guess for $A$), how might you do that? In our paper with \mdsine{} we did this in a two-step process. First we estimated the conditional expectation  $\bar b_{ij}=\mathbb E (b_{ij} \mid z_{ij}=1)$ by computing the conditional means of $b_{ij}$ where we only included those Gibbs samples for when the corresponding $z_{ij}$ was a 1 (every $i,j$ pair will have a different index of Gibbs samples to include in this calculation).  Then our best estimate for $A$ was defined as 
\begin{equation*}
a_{ij}=\begin{cases}
    \bar b_{ij} & \text{if } \mathbb P(z_{ij}=1)>0.5 \\
    0 & \text{otherwise}
\end{cases}
\end{equation*} where $\mathbb P(z_{ij}=1)$ is estimated by counting the number of Gibbs samples where $z_{ij}=1$  and dividing by the total number of Gibbs sample.

What was performed in \mbt{} was to approximate $A$ by taking the mean of $B$ over all the Gibbs samples. This is problematic because for a Gibbs sample where $z_{ij}=0$, the corresponding value for $b_{ij}$ is sampled from the prior (nothing is being learned with those samples and they, by construction, contribute nothing to the dynamics).  To circumvent these kinds of issues when deciding what might be the best possible parameter estimate given the complicated structure of our model (including module learning) we opted for the forecasting as described in the previous paragraph. Simply make a forecast for each Gibbs sample and consider those forecasts as a distribution and take the median value of the taxa abundance for each time-point as our best estimate.

\end{appendix}
\end{document}

%% file: macros.tex
\renewcommand{\phi}{\varphi}
\renewcommand{\epsilon}{\varepsilon}

\newcommand{\appropto}{\mathrel{\vcenter{
  \offinterlineskip\halign{\hfil$##$\cr
    \propto\cr\noalign{\kern2pt}\sim\cr\noalign{\kern-2pt}}}}}